\documentstyle[art10_ikf,hip-100a_ikf,epsf]{article}
\begin{document}

\begin{center}
{\bf INSTITUT~F\"{U}R~KERNPHYSIK,~UNIVERSIT\"{A}T~FRANKFURT}\\
60486 Frankfurt, August--Euler--Strasse 6, Germany
\end{center}

\hfill IKF--HENPG/2--96
\vspace{2cm}

\begin{center}
{  \Large \bf  Pion and Strangeness Puzzles
}
\end{center}


\vspace{0.3cm}
\begin{center}
Marek Ga\'zdzicki\footnote{E--mail address:
marek@ikf.physik.uni--frankfurt.de}
\end{center}

\vspace{0.3cm}
Institut f\"ur Kernphysik, Universit\"at Frankfurt, 
Frankfurt, Germany
\vspace{1cm}

\begin{center}
{\bf Abstract}
\end{center}
\noindent
Dependence of pion and strangeness production on number
of participant nucleons and collision energy is discussed for
central A+A collisions.
A possible interpretation of the experimental
results  assuming  transition to
QGP is sketched within a simple statistical approach.

\vspace{3cm}
\noindent
{\it Talk given at STRANGENESS'96, Budapest, Hungary, May 15--17 1996} \\
{\it to be published in the special issue of Heavy Ion Physics}   

\vfill
\today

\newpage 
 
\section{Introduction}

Massive hadrons are effective degrees of freedom in strongly
interacting matter at low energy density. 
We expect that in the matter at high enough energy density  the
degrees of freedom are almost massless quarks and gluons.
This new form of matter conjectured long ago \cite{Co:75}
is called Quark Gluon Plasma (QGP) \cite{Sh:80}. 
By studies of the transition to QGP and the QGP properties itself
we can better understand the structure of the QCD vaccum, the
origin of the hadron masses and confinement of quarks inside
hadrons.

Relativistic nucleus--nucleus collisions (A+A) are used as a tool
to study strongly interacting matter under extreme conditions.
In the laboratory we can control two basic parameters determining the 
property of the matter: the size of the colliding nuclei and the collision
energy.
With the increasing size of the colliding nuclei we increase the volume
of the created matter and its life time.
Increasing collision energy we increase the initial energy density of the
created matter.
Thus the transition from hadronic matter (HM) to QGP
should be observed as a  change of the collision properties
studied as a function of the size of colliding nuclei and/or
collision energy.

The characteristic feature of the transition from hadronic matter to QGP
is an increase of the effective number of degrees of freedom 
and the reduction of their masses \cite{QGP2}.
The increase of the effective number of degrees of freedom should 
cause an increase of entropy production.
Thus the produced entropy, at high energies mainly determined by
the multiplicity of pions, can be considered as one of the important 
observables in search for a transition to QGP \cite{Va:82}.
The reduction of the effective masses of the degrees of freedom
should result in the weaker dependence of the ratio
{\it particle number/entropy} on the temperature of matter
(collision energy).
The corresponding observable is the ratio {\it strangeness/entropy}
\cite{Re:85,Ka:86},
at high energy mainly determined by the ratio {\it strangeness/pion};
by strangeness we mean here total number of $s$  and $\overline{s}$
quarks in the system.
Note that  entropy, strangeness and baryon number are defined in any form
of strongly interacting matter like QGP or HM.
Therefore their values (or ratios) can be followed through
hadronization and freeze--out stages
of the collision when the distribution of entropy, strangeness and baryon
number among final states hadrons takes place. 

We argue below that the existing data on pion and strangeness
production in central collisions of heavy nuclei
indicate rapid changes occuring between
BNL AGS ($\approx $ 15 A$\cdot$GeV/c) and CERN SPS 
($ \approx $ 200 A$\cdot$GeV/c)
collision energies.
Can these changes be interpreted as due to transition from 
HM to QGP?

This presentation reviews  several recent works
of Dieter R\"ohrich and myself \cite{Ga:95a,Ga:95b,Ga:96}
where the details of the analysis and 
the references to the original
experimental papers used in the compilations
\cite{Ga:95a,Ga:96} are given.

The paper is organized as follows. In the Section 2 the main
properties of the  analysis of the experimental
data are described.
In Sections 3 and 4 the experimental results are presented.
The dependence of pion and strangeness production  on  
the number of participant nucleons and collision energy is shown.
The results are discussed in Section 5.
The paper is closed by summary and conclusions given in Section 6.
 
\section{Analysis}

The aim of this paper is to study basic features of pion and strangeness
production in nuclear collisions as a function of the size
of the colliding nuclei and the collision energy.
This goal suggests a specific way of data selection and presentation.

Matter created in A+A collisions expands in a complicated way. 
This expansion depends on the size of the colliding nuclei and
collision energy.
In order to reduce the  influence of the expansion on the results
and allow comparison of the data for various systems and energies
we limit analysis to the data integrated over the full momentum space.

Comparison and
interpretation of the data can be further simplified by concentrating on 
the data for head--on collisions of identical nuclei.
Thus in the present paper the analysis is limited to the central
collisions of similar nuclei.

Even in head--on  collisions of identical nuclei not all nucleons
participate in the interaction. 
Therefore  to study the volume dependence 
an average number of participant nucleons, $\langle N_P \rangle$,
measured experimentaly for each reaction is used instead of
the total number of nucleons in the colliding nuclei.
The A+A results are compared with the corresponding results for
all inelastic nucleon--nucleon (N+N) interactions.
In the case of N+N interactions the number of participant nucleons
is taken to be~2.

The collision energy dependence is studied using the Fermi energy
variable \cite{Fe:50,La:53}:
\begin{equation}
F = \frac {(\sqrt{s}_{NN} - 2 m_N)^{3/4} } { \sqrt{s}_{NN}^{1/4} },
\end{equation} 
where $\sqrt{s}_{NN}$ is the c.m. energy for a nucleon--nucleon pair and
$m_N$ is the mass of the nucleon.
There are several advantages in using $F$ as an energy variable.
The measured mean pion multiplicity in N+N interactions, 
$\langle \pi \rangle_{NN}$,
in the studied collision energy range is approximately proportional
to $F$ \cite{Ga:95a,Ga:95b,Go:89}.
In the Landau model \cite{La:53} both the entropy and
the initial temperature of the matter (for
$\sqrt{s}_{NN} >> 2 m_N$) are also proportional to $F$.

\section{Pion Puzzle}
 
The dependence of the average pion multiplicity per participant
nucleon, $\langle \pi \rangle/\langle N_P \rangle$, on the number of 
participant nucleons is shown in Fig. 1 for three different collision
energies.

Results for central collisions of identical nuclei (A $>$ 30)
are plotted together with the results for N+N interactions.
The $\langle \pi \rangle/\langle N_P \rangle$ ratio
for central A+A collisions at all energies  is independent 
of $\langle N_P \rangle$ \cite{Ga:95c}.
At low energies (p$_{LAB} < $ 15 A$\cdot$GeV/c)
the value of the saturation level of the 
$\langle \pi \rangle/\langle N_P \rangle$
ratio for central A+A collisions
is significantly lower than the corresponding value for
N+N interactions.
This effect is called {\bf pion suppression} in low energy
A+A collisions.
At high energies {\bf pion enhancement} is observed;
the saturation level for central A+A collisions is higher
than the corresponding ratio for N+N interactions.

The independence of the $\langle \pi \rangle/\langle N_P \rangle$
ratio of $\langle N_P \rangle$ simplifies a study of the energy
dependence for which data for various nuclei  can be put together.
The difference between the ratio
$\langle \pi \rangle/\langle N_P \rangle$
for central A+A collisions and N+N interactions:
\begin{equation}
\Delta \frac {\langle \pi \rangle} {\langle N_P \rangle} =
\frac { \langle \pi \rangle_{AA} } { \langle N_P \rangle_{AA} } -
\frac { \langle \pi \rangle_{NN} } { \langle N_P \rangle_{NN} } 
\end{equation}
is plotted in Fig. 2 as a function of $F$.

The difference is energy independent and equal to about --0.35
for low energy collisions (p$_{LAB} < $ 15 A$\cdot$GeV/c).
The low energy scaling is violated by high energy results
(Pb+Pb at 158 A$\cdot$GeV/c and S+S at 200 A$\cdot$GeV/c),
where the enhancement of pion production is observed.
A possible origin of this unusual
energy behaviour, called {\bf pion puzzle},
is discussed in Section 5.

\section{Strangeness Puzzle}

Total production of strangeness relative to pion production
is studied using the ratio \cite{Al:94}:
\begin{equation}
E_S = \frac { \langle \Lambda \rangle + \langle K + 
\overline{K} \rangle }
{ \langle \pi \rangle },
\end{equation}
where $\langle \Lambda \rangle$ is the mean multiplicity of produced
$\Lambda/\Sigma^0$ hyperons and $\langle K + \overline{K} \rangle$
is the mean multiplicity of kaons and antikaons.
The dependence of the $E_S$ ratio on $\langle N_P \rangle$ is
shown in Fig. 3 for three different collision energies.
The results for central A+A collisions \cite{Ma:96} are shown together with
the results for N+N interactions.
Due to the fact that the data on strangeness production are sparse
the data for collisions between non--equal mass
nuclei are also included (C+Cu/Zr at 4.5 A$\cdot$GeV/c,
Si+Au/Pb at 14.6 A$\cdot$GeV/c \cite{Ei:92} and 
S+Ag at 200 A$\cdot$GeV/c).

There is a significant increase of the relative strangeness production
(measured by the $E_S$ ratio) when going from N+N interactions to
central A+A collisions at all studied collision energies.
This increase is called {\bf strangeness enhancement}.
It appears to be strongest at BNL AGS energy.
The relative strangeness production  saturates for large
enough values of $\langle N_P \rangle$ at AGS BNL and CERN SPS energies.
The saturation effect can not be established at 4.5 A$\cdot$GeV/c 
as the results for collisions of heavy nuclei do not exist at this
energy.

The collision energy dependence of the $E_S$ ratio is shown in Fig. 4.
The results for N+N interactions are shown in Fig. 4a.
A monotonic increase of $E_S$ between Dubna energy 
(p$_{LAB}$ = 4.5 A$\cdot$GeV/c) and CERN SPS energy 
(p$_{LAB}$ = 200 A$\cdot$GeV/c)
is observed.
In the range from 15 A$\cdot$GeV/c to 200 A$\cdot$GeV/c  the $E_S$ ratio
for N+N interactions increases by a factor of about 2.
A qualitatively different energy dependence  is
observed for central A+A collisions.
The rapid increase of the $E_S$ between Dubna and BNL AGS energies
is followed by a weak change of the $E_S$ between BNL AGS and CERN SPS
collision energies (Fig. 4b).

A possible interpretation of the {\bf strangeness puzzle} --
'strange' energy dependence of the relative strangeness
production in A+A collisions --
is discussed in the next section. 

\section{Discussion}

The saturation of the relative pion 
($\langle \pi \rangle/\langle N_P \rangle$) 
and strangeness ($E_S$)
production with the volume of the created system 
($\langle N_P \rangle$) may be treated as an indication that
both entropy and strangeness approached their equilibrium values.
Thus the interpretation of the results in terms of statistical
models is suggested by the data.

It can be argued \cite{La:53,Le:94,Ge:94}
that both entropy and strangeness
reach saturation at the early and hot 
stages of the collision and  their values are only weakly
affected by later stages of  system evolution.
Further interpretation of the present results is therefore
based on the simplifing assumption
that the measured final state entropy and strangeness
reflect the equilibrium values at high temperature stage.
Thus studing the entropy and strangeness production one can try to
deduce the properties of matter (HM or QGP) at the early stage of 
the collision.

Transition from HM to QGP causes  an increase
of the effective number of degrees of freedom resulting in
the increase of entropy production.
As pointed out by Landau  \cite{La:53}
the entropy of the system
is approximately proportional to the pion multiplicity \cite{He:96}.
The entropy contained in baryons and generated  during their thermalization 
process is not directly sensitive to the initial number of
degrees of freedom and therefore should not be counted. 
Thus the relevant quantity is the entropy contained in the produced
particles -- the {\it inelastic} entropy.
However during the expansion process a  fraction of the initially created
inelastic entropy can be transfered back to the baryonic environment.
The experimental observation of pion suppression at low
energies (see Figs. 1 and 2)
can be interpreted as a result of the inelastic entropy transfer 
to baryons.
Following arguments given in Ref. \cite{Ga:95b} one can estimate the 
inelastic
entropy as:
\begin{equation}
S \sim \langle \pi \rangle + \alpha \cdot \langle N_P \rangle,
\end{equation}
where $\alpha \cdot \langle N_P \rangle$ is a correction for the
inelastic entropy transfer to baryons ($\alpha \approx -0.35$).
For simplicity the contribution from other produced particles
(mainly kaons)
is neglected in Eq. 4, but it is included in the final evaluation of
$S$ presented in Fig. 5a \cite{Ga:95b}.

The unusual increase of the relative pion production at
CERN SPS energies may be interpreted as due to an increase of 
the effective number of degrees of freedom.
In fact,
in the generalized Landau model \cite{Ga:95b} the inelastic entropy
is proportional to:
\begin{equation}
S \sim g^{1/4} \cdot \langle N_P \rangle \cdot F,
\end{equation}
where $g$ is the effective number of degrees of freedom.
Thus the observed deviation of the data for A+A collisions 
from the Landau scaling:
\begin{equation}
\frac {S} {\langle N_P \rangle} \sim F
\end{equation}
can be interpretad as due to an increase of the effective number of 
degrees of freedom when crossing the transition collision energy. 
The magnitude of this increase can be estimated,
within the generalized Landau model, as forth power of the ratio of 
slopes of straight lines describing (see Fig. 5a) 
low and high energy A+A data:
1.33$^4$ $\approx$ 3 \cite{Ga:95b}.
A hypothetical dependence of $S/\langle N_P \rangle$ on $F$
is shown by solid lines in Fig. 5a;
the transition region is assumed to be at
about p$_{LAB}$ = 30 A$\cdot$GeV/c \cite{Hu:95}.

The second dominant effect of the transition from HM
to QGP is the reduction of the effective masses of degrees of
freedom.
Basic thermodynamics tells us that for massless particles 
the ratio {\it particle number/entropy}  is independent of temperature.
For massive particles the ratio increases with $T$ at low
temperature approaching saturation level (equal to the 
corresponding ratio for massless particles) at high temperatures,
$T >> m$.
This property can be used to study the magnitude of the 
effective mass
of strangeness carriers in strongly interacting matter.
The $E_S$ ratio is approximately proportional to the ratio 
{ \it number of strangeness carriers/entropy (strangeness/entropy)} 
and therefore
its temperature (collision energy, $F$) dependence should be sensitive
to the effective mass of strangeness carriers. 
Reduction of the mass of
strangeness carriers should cause a weaker dependence of
the $E_S$ ratio on the collision energy.   

The rapid increase of the $E_S$ ratio in the energy 
range of $ F < $ 2 GeV$^{1/2}$ (see Fig. 4b)
can be interpreted as due to large effective mass of strangeness
carriers
(kaons or constituent strange quarks, $ m_S \approx $ 500 MeV )
in comparison to the temperature of matter, $T < T_C \approx $ 150 MeV.
At temperature above $T_C$ matter is in the form of QGP
and mass of strangeness carriers is equal to the mass of
current strange quarks, $ m_S \approx $ 150 MeV, consequently
$ m_S \leq T $.
Thus much weaker dependence of the $E_S$ ratio on $F$ is expected 
in the high energy region where the creation of QGP takes place.

The equilibrium value
of the {\it strangeness/entropy} ratio
is higher in HM than in QGP at very high temperatures.
This is due to the fact that it is proportional
to the ratio of the effective number of strangeness degrees of freedom
to the number of all degrees of freedom.
This ratio is lower in QGP than in hadronic matter.
At low temperature, however, the {\it strangeness/entropy}
ratio is lower in HM than in QGP.
This is caused, as previously discussed, by the high mass of strangeness
carriers in HM. 
Thus, in general, a transition to QGP may lead to an increase or a decrease
of the {\it strangeness/entropy} ratio depending at which temperatures
of QGP and HM the  comparison is made.
A quantative estimation of the {\it strangeness/entropy} dependence
on temperature within simple models of QGP and HM can be found
in Ref. \cite{Ka:86}.
The estimated temperature of the crossover of QGP and HM dependences
is about 130 MeV \cite{Ka:86}.
The measured low values of  the $E_S$ ratio at CERN SPS
relative to the values at BNL AGS can be interpreted as due to
reduction of the $E_S$ value at the transition to QGP. 
The hypothetical dependence of the $E_S$ ratio on $F$ is indicated
by the solid lines in Fig. 5b.
The transition region is assumed to be at 
about p$_{LAB}$ = 30 A$\cdot$GeV/c \cite{Hu:95}.

\section{Summary and Conclusions}
 
The experimenal data on pion and strangeness production indicate:\\
(a) saturation of pion and strangeness production with the
number of participant nucleons,\\
(b) change in the collision energy dependence occuring between
15 A$\cdot$GeV/c and 200 A$\cdot$GeV/c.\\
\noindent
Within a simple statistical approach
the observed behaviour can be qualitatively understood as due to:\\
equilibration of entropy  and strangeness production in
collisions of heavy nuclei (a),\\
transition from hadronic matter to QGP occuring between
BNL AGS and CERN SPS energies (b).\\
\noindent
These observations hold already for the central S+S collisions,
they are not unique to central Pb+Pb collisions.

A quantitative description of the existing data
within  theoretical models  and
experimental study of the energy dependence of pion and strangeness
production between BNL AGS and CERN SPS energies are needed
for a final clarification of the observed pion and strangeness
puzzles.
 
\begin{center}
{\bf Acknowledgments}
\end{center}
I would like to thank Reinhard Stock for numerous discussions and
comments .

\vfill\eject 

\newpage
\vspace*{16cm}
\includegraphics{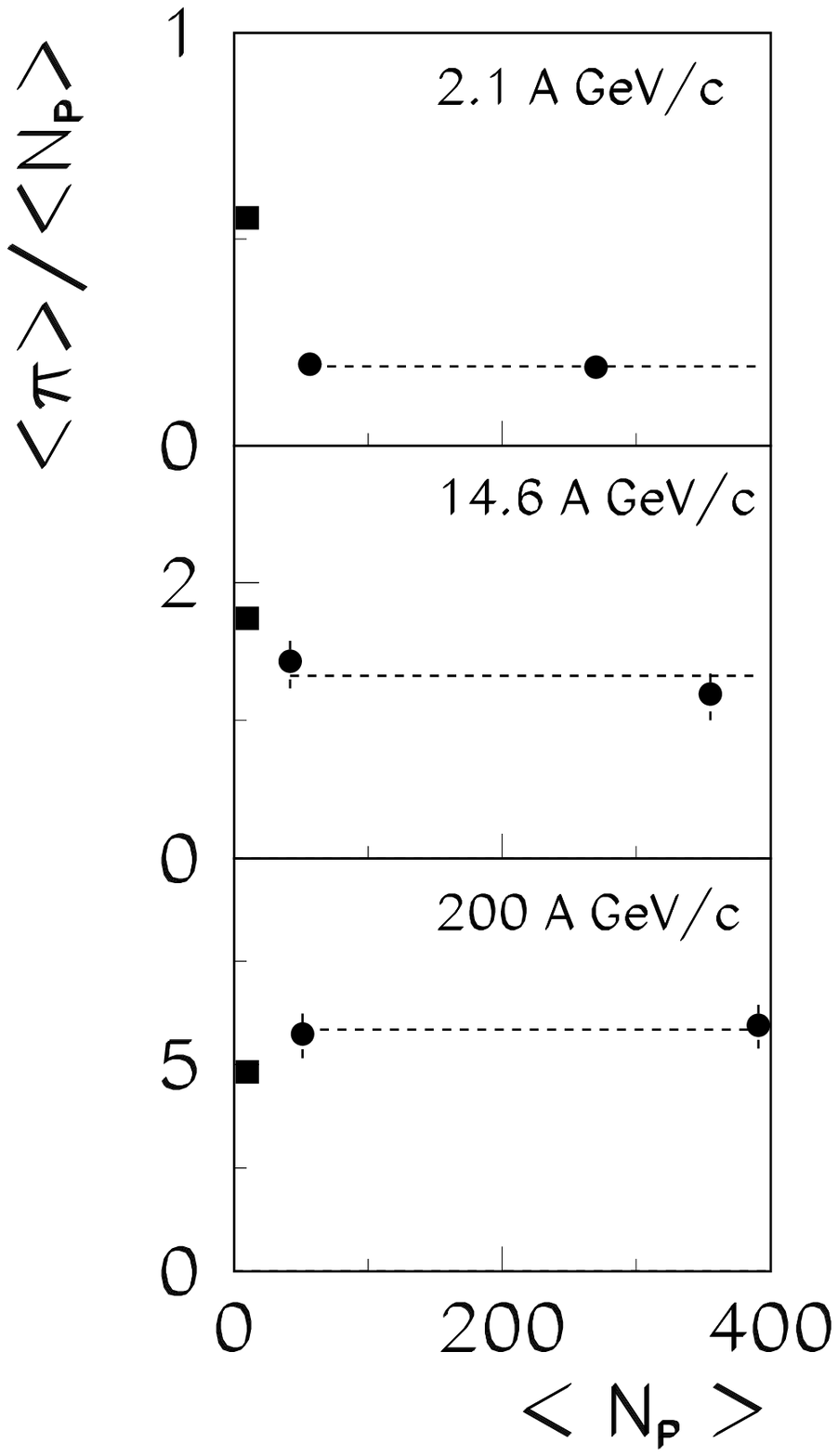}
\vskip -70pt
\begin{minipage}[t]{12cm}
\noindent \bf Fig. 1.  \rm
The dependence of the ratio $\langle \pi \rangle/\langle N_P \rangle$
on $\langle N_P \rangle$ at three different collisions energies
(p$_{LAB}$ = 2.1, 14.6  and 200 A$\cdot$GeV/c).
The data for Au+Au at 11.6 A$\cdot$GeV/c and Pb+Pb at 158 A$\cdot$GeV/c
were extrapolated to 14.6 A$\cdot$GeV/c and 200 A$\cdot$GeV/c,
respectively.
The data for central A+A collisions are indicated by circles and
the data for N+N interactions by squares.
\end{minipage}
\vskip 4truemm

\newpage
\vspace*{16cm}
\includegraphics{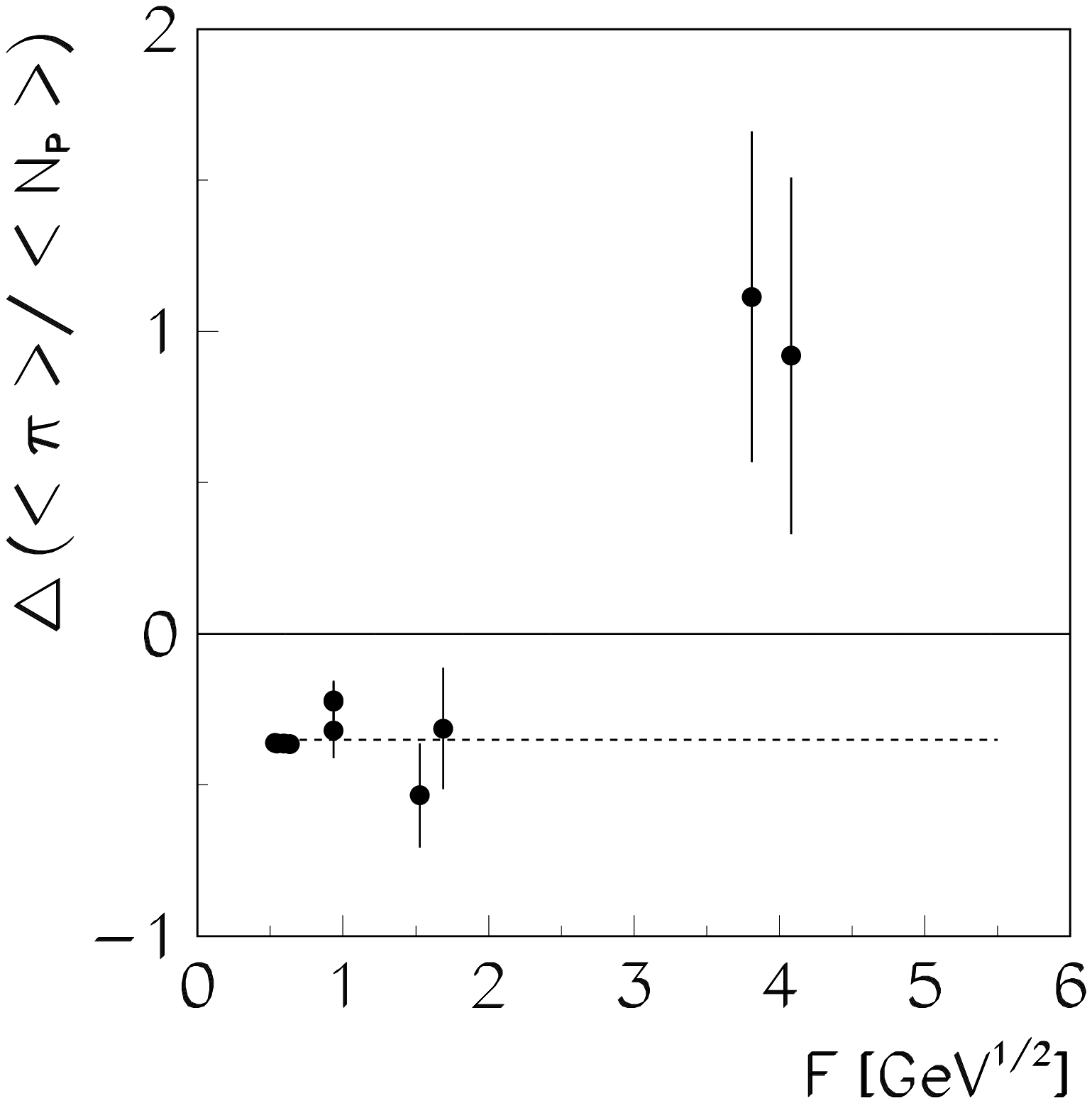}
\vskip -70pt
\begin{minipage}[t]{12cm}
\noindent \bf Fig. 2.  \rm
The dependence of the difference 
$\Delta(\langle \pi \rangle/\langle N_P \rangle$) (see Eq. 2) on the
collision energy measured by the Fermi energy variable, $F$ (see
Eq. 1).
\end{minipage}
\vskip 4truemm

\newpage
\vspace*{16cm}
\includegraphics{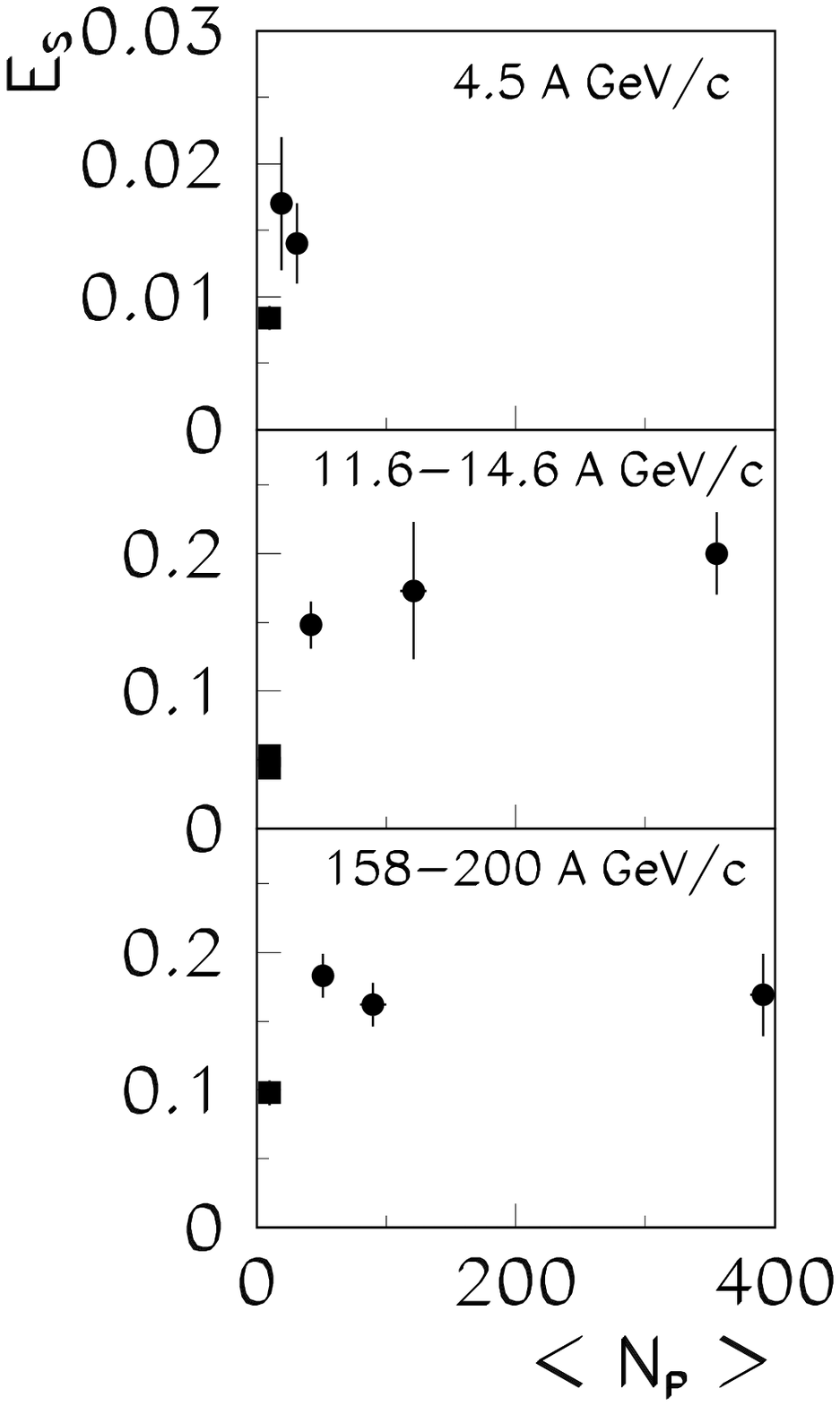}
\vskip -70pt
\begin{minipage}[t]{12cm}
\noindent \bf Fig. 3.  \rm
The dependence of the $E_S$ ratio (see Eq. 3)
on $\langle N_P \rangle$ at three different collisions energies
(p$_{LAB}$ = 4.5, 11.6--14.6 and 158--200 A$\cdot$GeV/c).
The data for central A+A collisions are indicated by circles and
the data for N+N interactions by squares.
\end{minipage}
\vskip 4truemm

\newpage
\vspace*{16cm}
\includegraphics{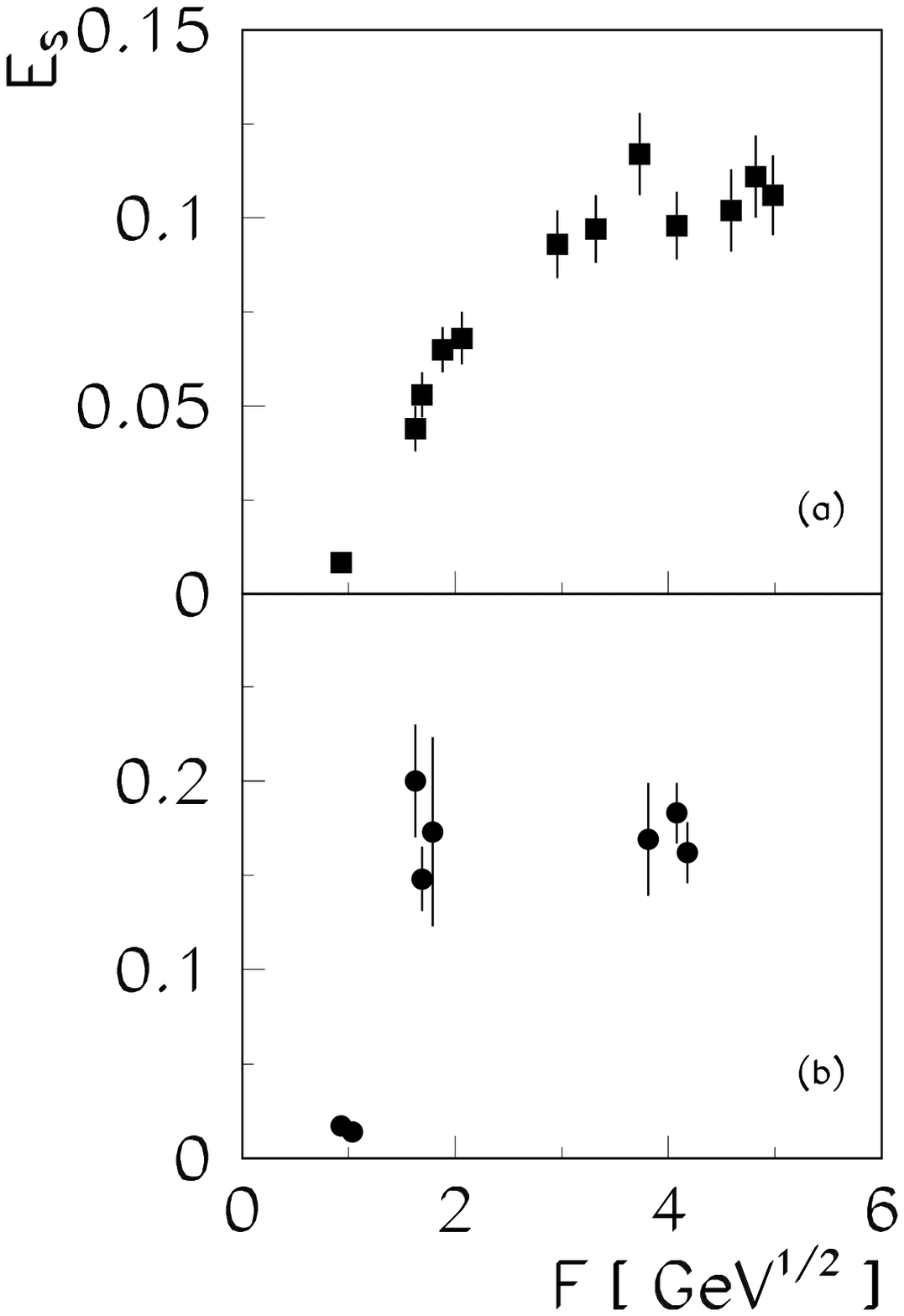}
\vskip -70pt
\begin{minipage}[t]{12cm}
\noindent \bf Fig. 4.  \rm
The dependence of the $E_S$ ratio 
(see Eq. 3) on the
collision energy measured by the Fermi energy variable, $F$ (see
Eq. 1)
for N+N interactions (a) and central A+A collisions (b).
\end{minipage}
\vskip 4truemm

\newpage
\vspace*{16cm}
\includegraphics{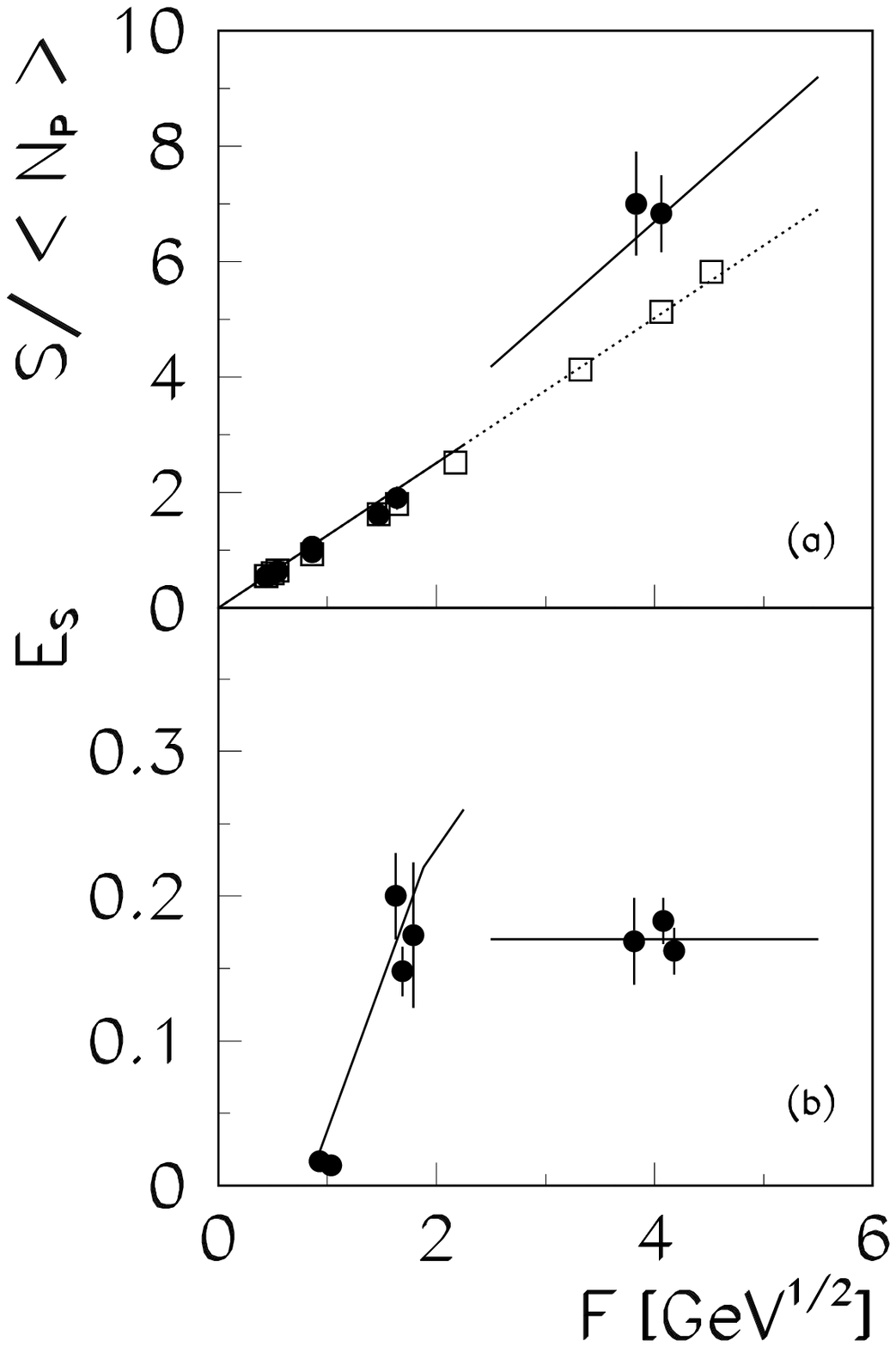}
\vskip -70pt
\begin{minipage}[t]{12cm}
\noindent \bf Fig. 5.  \rm
(a) 
The dependence of the $S/\langle N_P \rangle$ ratio 
(see Eq. 4) on the
collision energy measured by the Fermi energy variable, $F$ (see
Eq. 1). The entropy, $S$, is measured in units given by pion entropy at
freeze--out. A hypothetical dependence of $S/\langle N_P \rangle$
ratio on $F$ for central A+A collisions assuming transtion to QGP
at p$_{LAB}$ = 30 A$\cdot$GeV/c is shown by solid lines.
The data for central A+A collisions are shown by circles and the data
for N+N interactions by squares.
(b) 
A hypothetical dependence of the $E_S$ ratio on $F$ assuming
transition to QGP at p$_{LAB}$ = 30 A$\cdot$GeV/c is shown
by a solid lines. The data for central A+A collisions
are indicated by circles. 
\end{minipage}
\vskip 4truemm


\begin{thebibliography}{99}\parindent=8truemm
\itemsep -1mm



\bibitem{Co:75} J. C. Collins and M. J. Perry, Phys. Rev. Lett.
{\bf 34} (1975) 151.

\bibitem{Sh:80} E. V. Shuryak, Phys. Rep. {\bf C61} (1980)
71 and {\bf C115} (1984) 151.

\bibitem{QGP2} Quark Gluon Plasma 2, Ed. R. C. Hwa,
{\it World Scientific} (1995)

\bibitem{Va:82} L. Van Hove, Phys. Lett. {\bf B118} (1982) 138.

\bibitem{Re:85} K. Redlich, Z. Phys. {\bf C27} (1985) 633.

\bibitem{Ka:86} J. Kapusta and A. Mekjan, Phys. Rev. {\bf D33}
(1986) 1304.

\bibitem{Ga:95a} M. Ga\'zdzicki and D. R\"ohrich, Z. Phys. 
{\bf C65} (1995) 215.

\bibitem{Ga:95b} M. Ga\'zdzicki, Z. Phys. {\bf C66} (1995) 659.

\bibitem{Ga:96} M. Ga\'zdzicki and D. R\"ohrich,
Z. Phys. {\bf C71} (1996) 55.

\bibitem{Fe:50} E. Fermi, Prog. Theor. Phys. {\bf 5} (1950) 570.

\bibitem{La:53} L. D. Landau, Izv. Akad. Nauk SSSR, Ser. Fiz.
{\bf 17} (1953) 51,\\
S. Z. Belenkij and L. D. Landau, Uspekhi Fiz. Nauk {\bf 56}
(1955) 309.

\bibitem{Go:89} A. I. Golokhvastov, Dubna Report, JINR E2--89--364
(1989).

\bibitem{Ga:95c} The Pb+Pb data at 158 A$\cdot$GeV/c are taken
from:\\
M. Ga\'zdzicki et al. (NA49 Collab.), Nucl. Phys. {\bf A590}
(1995) 197c.

\bibitem{Al:94} T. Alber et al. (NA35 Collab.),
Z. Phys. {\bf C64} (1994) 195.

\bibitem{Ei:92} S. E. Eiseman et al. (E810 Collab.),
Phys. Lett. {\bf B297} (1992) 44, \\
T. Abbott et al. (E802 Collab.), Phys. Rev. {\bf C50}
(1994) 1024. 

\bibitem{Ma:96} The $E_S$ ratio for Pb+Pb collisions at 
158 A$\cdot$GeV/c was estimated on the basis of the preliminary
NA49 results for kaon production presented at this Conference
by S. Margetis.

\bibitem{Le:94} J. Letessier, A. Tounsi and J. Rafelski,
Phys. Rev. {\bf C50} (1994) 406, \\
J. Letessier, A. Tounsi and J. Rafelski,
Acta Phys. Pol. {\bf 27} (1996) 1035 and references therein.

\bibitem{Ge:94} K. Geiger, Nucl. Phys. {\bf A566} (1994) 257c and
references therein.

\bibitem{He:96} for detailed discussion of the influence of
non-equilibrium processes at freeze--out see::\\
J. Sollfrank and U. Heinz, Phys. Lett. {\bf 289} (1992) 132, \\
S. Ochs and U. Heinz, University of Regensburg Report, TPR 96--09,
hep--ph/9606458 (1996).

\bibitem{Hu:95} C. M. Hung and E. V. Shuryak, Phys. Rev. Lett.
{\bf 75} (1995) 4007
















%




 










\end{thebibliography}
\end{document}